\documentclass[review,12pt]{elsarticle}
\usepackage{amssymb}
\usepackage{here}
\usepackage{hyperref}

\journal{Nuclear Inst. and Methods in Physics Research A}

\begin{document}

\begin{frontmatter}

\title{Injection and capture of antiprotons in a Penning-Malmberg trap using a drift tube accelerator and degrader foil}
\author[1]{C.~Amsler}
\author[2]{H.~Breuker}
\author[14,11]{M.~Bumbar}
\author[1]{S.~Chesnevskaya}
\author[3,4]{G.~Costantini}
\author[5,6]{R.~Ferragut}
\author[6]{M.~Giammarchi}
\author[1]{A.~Gligorova}
\author[3,4]{G.~Gosta}
\author[7]{H.~Higaki}
\author[8,9,10]{M.~Hori}
\author[1]{E.D.~Hunter\fnref{l5}}
\fntext[l5]{Present address: Experimental Physics Department, CERN, 1211 Geneva, Switzerland}
\author[1]{C.~Killian}
\author[1,11]{V.~Kraxberger}
\author[12]{N.~Kuroda\corref{cor1}}
\ead{kuroda@phys.c.u-tokyo.ac.jp}
\author[1,11]{A.~Lanz\fnref{l2}}
\fntext[l2]{Present address: University College London, Gower St, London WC1E 6BT, United Kingdom}
\author[3,4]{M.~Leali}
\author[6,13]{G.~Maero}
\author[14]{C.~Malbrunot\fnref{l1}}
\fntext[l1]{Permanent address: TRIUMF, Vancouver, BC V6T 2A3, Canada}
\author[3,4]{V.~Mascagna}
\author[12]{Y.~Matsuda}
\author[1]{V.~M\"{a}ckel\fnref{l3}}
\fntext[l3]{Present address: INFICON GmbH, Bonner Str. 498, 50968 K\"{o}ln, Germany}
\author[3,4]{S.~Migliorati}
\author[1]{D.J.~Murtagh}
\author[15]{Y.~Nagata}
\author[1,11]{A.~Nanda}
\author[14,11]{L.~Nowak}
\author[6,13]{M.~Rom\'e}
\author[1]{M.C.~Simon}
\author[16,17]{M.~Tajima\corref{cor1}}
\ead{minori.tajima@spring8.or.jp}
\author[5,6]{V.~Toso}
\author[2]{S.~Ulmer}
\author[3,4]{L.~Venturelli}
\author[1,11]{A.~Weiser}
\author[1]{E.~Widmann}
\author[2]{Y.~Yamazaki}
\address[1]{Stefan Meyer Institute for Subatomic Physics, Austrian Academy of Sciences, Dominikanerbastei 16, 1010 Vienna, Austria}
\address[2]{Ulmer Fundamental Symmetries Laboratory, RIKEN, 2-1 Hirosawa, Wako, 351-0198 Saitama, Japan}
\address[14]{Experimental Physics Department, CERN, 1211 Geneva, Switzerland}
\address[11]{Vienna Doctoral School in Physics, University of Vienna, Universit\"atsring 1, 1010 Vienna, Austria}
\address[3]{Dipartimento di Ingegneria dell'Informazione, Universit\`{a} degli Studi di Brescia, via Branze 38, 25123 Brescia, Italy}
\address[4]{INFN~-~Sezione di Pavia, via Agostino Bassi 6, 27100 Pavia, Italy}
\address[5]{L-NESS and Department of Physics, Politecnico di Milano, via Anzani 42, 22100 Como, Italy}
\address[6]{INFN~-~Sezione di Milano, via Giovanni Celoria 16, 20133 Milan, Italy}
\address[7]{Graduate School of Advanced Science and Engineering, Hiroshima University, 1-4-1, Kagamiyama, Higashihiroshima, 739-8527 Hiroshima, Japan}
\address[8]{Blackett Laboratory, Imperial College London, London SW7 2AZ, United Kingdom}
\address[9]{QUANTUM, Institut f\"ur Physik, Johannes Gutenberg-Universit\"at Mainz, 55128 Mainz, Germany}
\address[10]{Max-Planck-Institut f\"{u}r Quantenoptik, Hans-Kopfermann-Stra{\ss}e 1, 85748 Garching, Germany}
\address[12]{Institute of Physics, Graduate School of Arts and Sciences, The University of Tokyo, 3-8-1 Komaba, Meguro-ku, 153-8902 Tokyo, Japan}
\address[13]{Dipartimento di Fisica, Universit\`{a} degli Studi di Milano, via Giovanni Celoria 16, 20133 Milan, Italy}
\address[15]{Department of Physics, Tokyo University of Science, 1-3 Kagurazaka, Shinjuku-ku, 162-8601 Tokyo, Japan}
\address[16]{RIKEN Nishina Center for Accelerator-Based Science, 2-1 Hirosawa, Wako, 351-0198 Saitama, Japan}
\address[17]{Japan Synchrotron Radiation Research Institute, 1-1-1 Kouto, Sayo-cho, Sayo-gun, 679-5198 Hyogo, Japan}
\cortext[cor1]{Corresponding author}

\begin{abstract}
The Antiproton Decelerator (AD) at CERN provides antiproton bunches with a kinetic energy of 5.3~MeV. 
The Extra-Low ENergy Antiproton ring at CERN, commissioned at the AD in 2018, 
now supplies a bunch of electron-cooled antiprotons at a fixed energy of 100~keV.
The MUSASHI antiproton trap was upgraded by replacing the radio-frequency quadrupole decelerator
with a pulsed drift tube to re-accelerate antiprotons and optimize the injection energy into the degrader foils. 
By increasing the beam energy to 119~keV, a cooled antiproton accumulation efficiency of $(26\pm6)$\% was achieved. 
\end{abstract}


\end{frontmatter}

\section{Introduction}
Fundamental research activities using low-energy antiprotons have been performed at the CERN 
Antiproton Decelerator (AD)~\cite{Maury1997} and Extra-Low ENergy Antiproton ring (ELENA)~\cite{Bartmann2018} 
for spectroscopic studies of exotic atoms 
such as antihydrogen~\cite{Ahmadi2017, Ahmadi2018, ALPHA2020, Eberhard2019, Gabrielse2012} 
and other antiprotonic atoms~\cite{Hori2016}, measurements of antiproton properties~\cite{Borchert2022, DiSciacca2013}, 
testing the weak equivalence principle~\cite{Kellerbauer2008, Perez2015, Bertsche2018, Anderson2023}, 
and studies of antiproton--matter interaction~\cite{Kirchner2011, Amsler2019, Aumann2022, Aghai2018, Aghai2021}.

The ASACUSA collaboration developed a slow antiproton beam source, the Monoenergetic UltraSlow Antiproton Source 
for High-precision Investigation (MUSASHI)~\cite{Kuroda2012}, 
to perform ground-state hyperfine spectroscopy of antihydrogen and other interaction studies 
with the combination of the former radio-frequency quadrupole decelerator (RFQD)~\cite{Lombardi2001} and the AD ring. 
The production of antihydrogen atoms in the so-called Cusp trap~\cite{Mohri2003} with beams from MUSASHI 
has been demonstrated~\cite{Enomoto2010}. 
They were extracted to a field-free region downstream of the positron-antiproton recombination 
location~\cite{Kuroda2014}, and their atomic quantum number distribution was measured~\cite{Kolbinger2021}. 
Since then, we have been working on increasing the antihydrogen beam intensity 
for high-precision spectroscopy by reducing the plasma temperature~\cite{Amsler2022}. 
A measurement of annihilation fragments from antiproton-nucleus scattering using ultraslow antiprotons from MUSASHI 
and thin targets was performed, and further studies are planned~\cite{Amsler2019}.

Before the commissioning of ELENA, a bunch of $3\times10^7$ antiprotons supplied from the AD at 5.3~MeV
was decelerated by the RFQD down to 120~keV which was found to be optimal, 
and injected into the Penning-Malmberg type MUSASHI trap.
The antiprotons were further decelerated to less than 10~keV 
by degrader foils at the entrance of the MUSASHI trap.
The foils maintained an ultra-high vacuum within the trap of $\le10^{-10}$~Pa, compared to $10^{-7}$~Pa in the RFQD,
and also acted as profile monitors for the injected antiprotons by detection of secondary electrons 
from thin silver strips printed on the surface of the foils~\cite{Hori2004}.
The ultra-high vacuum is essential to suppress antiproton annihilations with the residual gas 
and to achieve efficient trapping, especially for stacking multiple antiproton bunches.
The antiproton injection point was steered by monitoring the beam profile on the foils, as described in Section~2.
Precise antiproton injection on the axis of the MUSASHI trap is required 
for stable confinement of non-neutral plasmas.
Otherwise, off-axis injection causes plasma instability such as Diocotron instability, resulting in antiproton loss.
The MUSASHI trap previously trapped up to $10^7$ antiprotons 
from 7 antiproton bunches with the RFQD at the AD facility~\cite{Kuroda2012}.

ELENA, commissioned at the AD in 2018, supplies a bunch of antiprotons at fixed 100~keV energy simultaneously 
to 4 users.
The first beam from ELENA to ASACUSA was delivered in 2021.
A thinner foil than the current degrader foils is one option to decelerate antiprotons from ELENA.
In this case, a mechanism is needed to optimize the energy of the emitted antiprotons from the foil
since the energy of the antiprotons from ELENA is fixed at 100~keV.
A remotely controlled degrader holder of thin foils with different thicknesses could be used.
It is however not easy to adjust the foil thickness with a hundred nanometer-scale precision, 
which is needed to optimize the antiproton energy in these experiments, 
and the thin foils can be damaged by the movable mechanism~\cite{Todoroki2016}. 

An accelerating drift tube is another option.
A pulsed drift tube to accelerate ion bunches has been used in many nuclear physics facilities 
after buffer-gas cooling of high-energy particles, see e.g.~\cite{Traykov2008, Schury2017}.
Since antiprotons annihilate in the buffer-gas, making this type of cooling unfeasible, a pulsed drift tube 
to directly decelerate antiprotons from 100~keV to a few keV was developed by GBAR~\cite{Husson2016} 
and PUMA~\cite{Fischer2024}.
Electrostatic optics for good focusing and insulating components to avoid discharges were carefully designed,
which is important for high-voltage switching at 100~kV.
The resulting energy is adjustable by changing the switching voltage of the drift tube.

We decided to use a pulsed drift tube to adjust the energy by up to +20~keV, as used previously with the RFQD,
and to decelerate the antiprotons using the current degrader foils.
This is an efficient solution for ASACUSA to work with ELENA 
since the discharge problem is less serious with 20~kV and we could keep the current degrader foils
without modifications.
Therefore, the drift tube accelerator was constructed, replacing the RFQD, and commissioned with ELENA.
The magnet, Penning trap, foil detector, and antiproton annihilation detectors used in this work 
have been described elsewhere~\cite{Kuroda2012}. 
In this paper we restrict ourselves to describing the upgrades and modifications made to the apparatus 
to allow its operation with the 100~keV antiproton beam from ELENA.

\begin{figure}[htb]
\includegraphics[width=14cm]{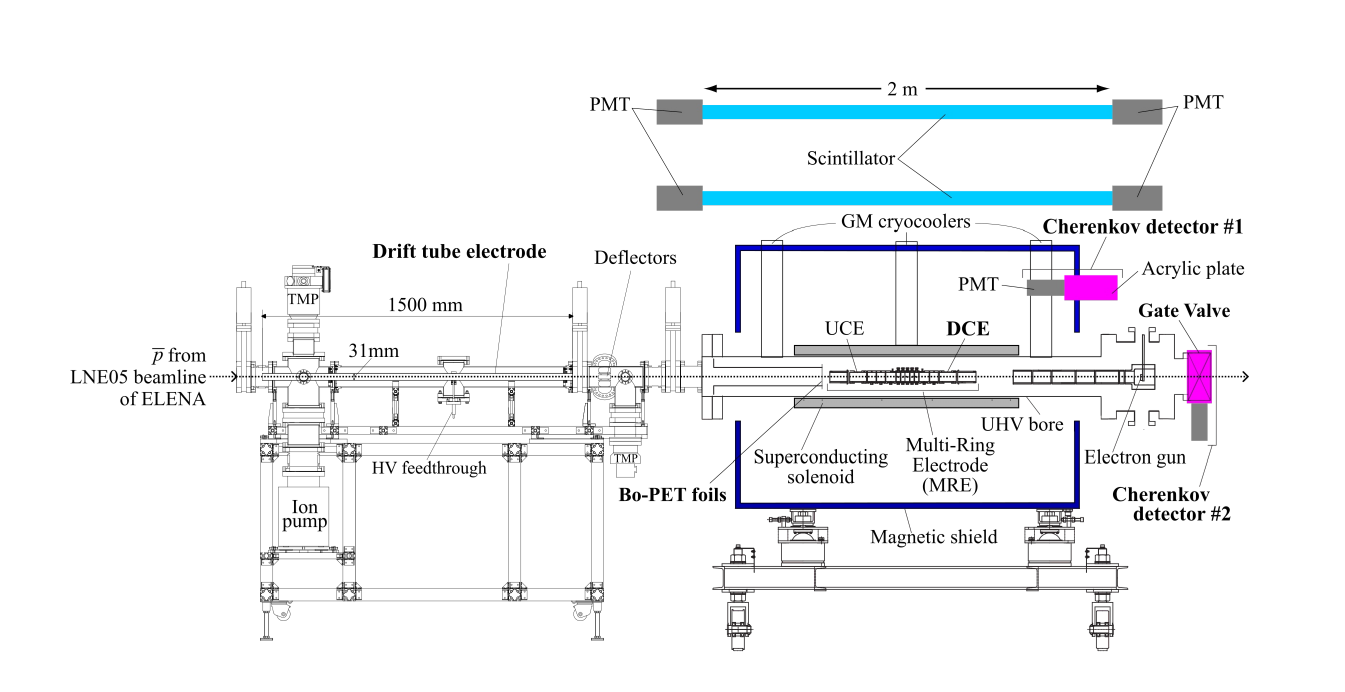}
\caption{
Experimental setup including the drift tube accelerator and the MUSASHI trap. 
A bunch of antiprotons is injected from the left.}
\label{exsetup}
\end{figure}

\section{Experimental setup}
Figure~\ref{exsetup} shows the experimental setup. 
The drift tube accelerator is connected to the existing MUSASHI trap detailed in~\cite{Kuroda2012}.
The MUSASHI trap is mainly composed of a superconducting solenoid of 2.5~T and 
Multi-ring electrodes (MRE)~\cite{Kuroda2005,Mohri1998} in a UHV bore.
The degrader foils at the entrance of the MUSASHI trap isolate the vacuum of $10^{-7}$ Pa inside the drift tube.
The foils are double-layered and made of biaxially oriented polyethylene terephthalate (Bo-PET) 
with a mass thickness of 90~$\mu$g~cm$^{-2}$ each.
Silver strips of 25~nm thickness deposited onto the surface of each foil act as anodes
readout by a peak-sensing ADC (CAEN, V785) or a digitiser (NI, PXI-6224),
see~\cite{Hori2004} for details.
There is a retractable electron gun downstream,
which is used to supply electrons prepared for antiproton cooling.
A gate valve, located at the downstream end, is closed and acts as a beam dump during the measurements of time of flight 
and the energy distribution of antiprotons.
The Cherenkov detector is composed of an acrylic plate (Mitsubishi Rayon, Acrylite-000) 
with a refractive index $n=1.49$.
Charged particles from annihilation (mainly pions) produce Cherenkov light 
in the acrylic plate, which is read out by a fine-mesh photo multiplier tube (Hamamatsu Photonics, R5505GX-ASSY),
see~\cite{Hori2003} for details.
One is attached to the downstream edge of the magnetic shield of the MUSASHI (detector \#1) and the other is attached to the side of the gate valve (detector \#2).
Two scintillator bars ($4\times6\times200$~cm$^3$) are located parallelly to the axis of the MUSASHI trap
with photomultiplier tubes (PMT Hamamatsu Photonics, H1949-50) connected to both ends of each bar. 
The distances between the trap axis and the scintillator bars are 82~cm and 150~cm, respectively.
A Time-to-Digital Converter (Agilent, U1051A) is used to record coincidence events of four PMTs.
The threshold of the energy deposit is set at approximately 5~MeV so that the event
is mainly due to charged pions according to GEANT4~\cite{Agostinelli2003} simulation,
see~\cite{Kuroda2005} for details.

The beam energy adjuster of the drift tube consists of a 1.5~m long electrode made of aluminum alloy 
with an inner diameter of 31~mm which is large enough compared with the injection beam size.
It is mounted within a stainless steel chamber with an inner diameter of 101~mm.
The magnetic field in the drift tube is less than 1~mT everywhere. 
The tube length of 1.5~m is designed to cover 95\% of the injected bunch length from ELENA with $4\sigma_{\rm rms}$ 
of 300~ns, $\delta p/p$ of $2.5\times10^{-3}$, and emittance (h/v) of $6/4$~$\pi\mu$m~\cite{Bartmann2018}.
The upstream side of the chamber is first evacuated by a turbomolecular pump (Shimadzu, TMP-303), 
and then isolated by a gate valve, to be further evacuated by an ionization pump (Varian, VacIon Plus 300 Triode).
A typical pressure of $6\times10^{-8}$~Pa is achieved, 
sufficiently low to connect to the upstream LNE05 beamline of $2\times10^{-8}$~Pa pressure.
A set of four electrostatic deflectors (rectangular plates) are installed in the downstream side of the vacuum chamber to optimize the beam direction
after acceleration by the drift tube.
The acceleration voltage is applied via a high-voltage feedthrough (CeramTec, 21144-01-CF) 
at the center of the long chamber of the drift tube, connected to a custom-made control circuit.

\begin{figure}[htb]
\centering
\includegraphics[clip,width=13cm]{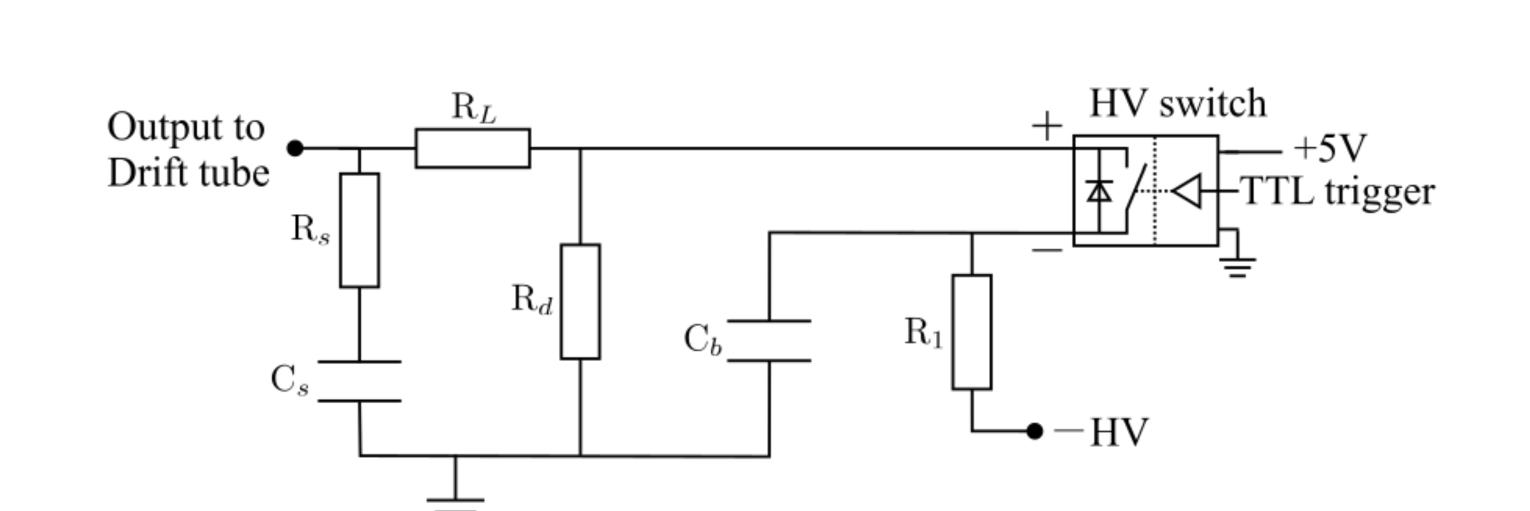}
\caption{
Diagram of the control circuit for the drift tube accelerator.
A negative HV power supply charges the capacitor bank (C$_b$).
A HV switch made of silicon carbide is used for fast switching triggered by external TTL input.
R$_d$ is adjusted to suppress DC offset voltage of the drift tube
caused by the leak current of the switch.
A snubber circuit composed of R$_s$ and C$_s$ is included in order to suppress voltage surges
when the switch is triggered, which could damage the switch by a reversed current (see also Tab.~\ref{tabc}).
}
\label{circuit}
\end{figure}

\begin{table}[htb]
\caption{Specifications of circuit components.}
\label{tabc}
\centering
\scalebox{0.6}{
\begin{tabular}{c|c|c}
\hline
\small{Symbol} & \small{Value} & \small{Description} \\
\hline \hline
C$_b$ & 10.5~nF & \small{three TDK, FHV-6AN} \\
\hline
R$_1$ & 1~M$\Omega$ & \small{KOA, GS 7LC 105K} \\
\hline
R$_L$ & 100~$\Omega$ & \small{KOA, PN-1 M F 101 J} \\
\hline
R$_d$ & 7.7~M$\Omega$ & \small{three KOA, GS 12LC 107K (100~M$\Omega$) and} \\
& & \small{one Ohmite, MOX95021005FVE (10~M$\Omega$)} \\
& & \small{in parallel} \\
\hline
R$_s$ & 200~$\Omega$ & \small{KOA, PN-1 M F 101 and two PN-0.5 C F 500 J} \\
\hline
C$_s$ & 50~pF & \small{Comet PCT, MINI-Cap CFMN-50EAC/35-DH-G} \\
\hline
\end{tabular}}
\end{table}

\begin{figure}[htb]
\centering
\includegraphics[clip,width=10cm]{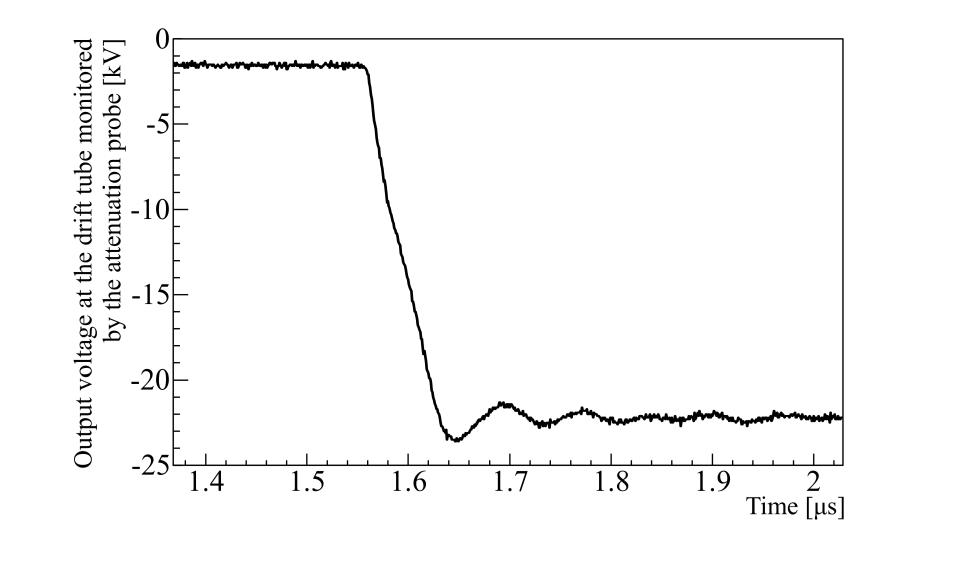}
\caption{
Time structure of the applied acceleration voltage monitored by an attenuation probe.
It takes 340~ns for antiprotons at 100~keV to go through the drift tube electrode.
}
\label{oscillo_pulse}
\end{figure}

\begin{figure}[htb]
\centering
\includegraphics[clip,width=7cm]{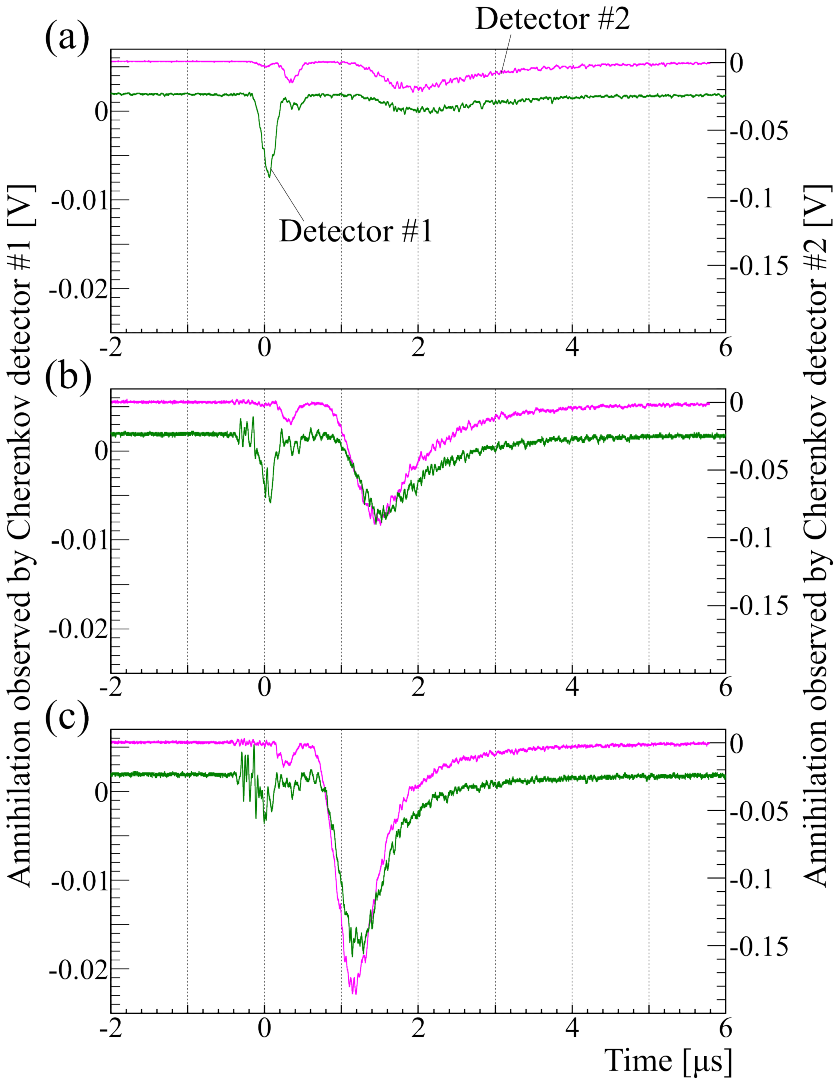}
\caption{
Output signals from the Cherenkov detectors when the acceleration voltage of the drift tube is 
(a)~0~kV, (b)~10~kV, and (c)~20~kV. 
The green and magenta lines correspond to detectors \#1 and \#2, respectively.
The broad peak at $t>0.6$~$\mu$s corresponds to antiprotons decelerated by the foils (with mean energy $\le 10$~keV).
With increasing acceleration voltage, more annihilations at the gate valve are observed at earlier times. 
This implies that the antiprotons are successfully accelerated by the drift tube. 
}
\label{cherenkov}
\end{figure}

Figure~\ref{circuit} shows a diagram of the control circuit (see also Table~\ref{tabc}).
It is designed to apply a HV pulse with an amplitude of $\le25$~kV and a fall time of 20~ns, 
assuming that the drift tube has an effective load inductance of 513~nH, 
capacitance of 85~pF, and resistance of 0.06~$\Omega$.
A negative high-voltage power supply (Matsusada, HGR30-30N) adjusted by an external DC control voltage 
is used to charge the capacitor bank (C$_b$, three TDK, FHV-6AN).
A commercial high-voltage switch made of silicon carbide (Behlke, HTS 301-60-SiC) is used for fast switching, 
triggered by external TTL input.
Three resistances R$_1$, R$_L$, and R$_d$ are selected not to exceed the current limit.
R$_d$ is adjusted to suppress DC offset voltage of the drift tube
caused by the leak current of the switch.
A snubber circuit composed of resistance R$_s$ and capacitance C$_s$ suppresses voltage surges 
and reverse currents when the switch is triggered, which can otherwise damage the switch.
The entire setup is covered by a cage of perforated metal sheets for safety. 
Simulations of this circuit show total switching time of 60~ns.
Figure~\ref{oscillo_pulse} shows the typical time structure of the applied pulse,
monitored by an attenuation probe (1/1000, Tektronix, P6015A).
The observed switching time is a bit longer than expected due to parasitic capacitance 
which are not included in the simulations.
It takes 340~ns for antiprotons at 100~keV to traverse the drift tube electrode.
Since the tube is grounded during the injection of the bunch, the bunch length is kept constant inside the tube.

\section{Test of the drift tube accelerator}
The antiproton bunch is detected by the Cherenkov detectors 
in synchronization with the injection trigger signal supplied by the ELENA facility.
Figure~\ref{cherenkov}(a)-(c) shows the PMT signals observed by the Cherenkov detectors 
for various acceleration voltages applied to the drift tube.
The green and magenta lines correspond to detectors \#1 and \#2, respectively.
The earlier sharp peak at $t=0$, mainly observed by detector \#1 which is closer to the degrader foils, 
corresponds to antiprotons accelerated by the drift tube (with a mean energy between 100 and 120~keV) annihilating in the foils.
The delayed broad peak at $t>0.6$~$\mu$s corresponds to antiprotons decelerated by the foils (with mean energy $\le 10$~keV).
They annihilate on the surface of the closed gate valve located 1.6~m downstream of the foils.
With increasing acceleration voltage, more annihilations at the gate valve are observed at earlier times. 
This implies that the antiprotons are successfully accelerated by the drift tube 
such that more antiprotons traverse the degrader foils and emerge from the downstream surface with a higher energy.
The small peak at $t=0.4$~$\mu$s, mainly observed by detector \#2 which is positioned at the gate valve, 
corresponds to fast antiprotons at 100~keV passing through some pin holes on the foils.

\begin{figure}[htb]
\centering
\includegraphics[clip,width=8cm]{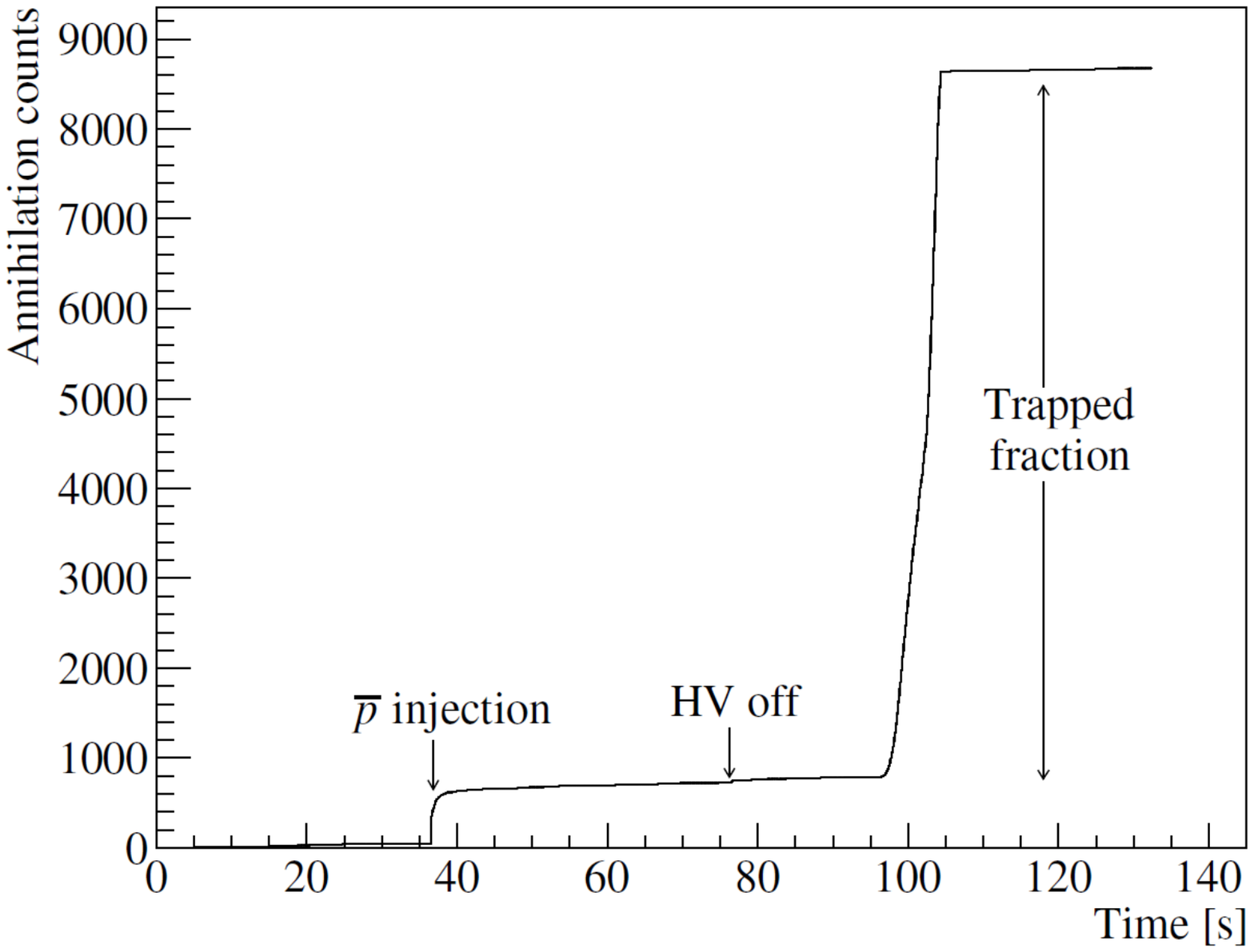}
\caption{
A typical time evolution of the cumulative annihilation counts measured by the two scintillator bars 
during one accumulation cycle.
Antiproton injection occurs at $t=36$~s, followed by electron cooling for 40~s, and slow extraction of the cooled antiprotons 
to the foils for 10~s after $t=95$~s.}
\label{TD_tstr}
\end{figure}

\section{Antiproton capture with ELENA}
\begin{figure}[htb]
\centering
\includegraphics[clip,width=8cm]{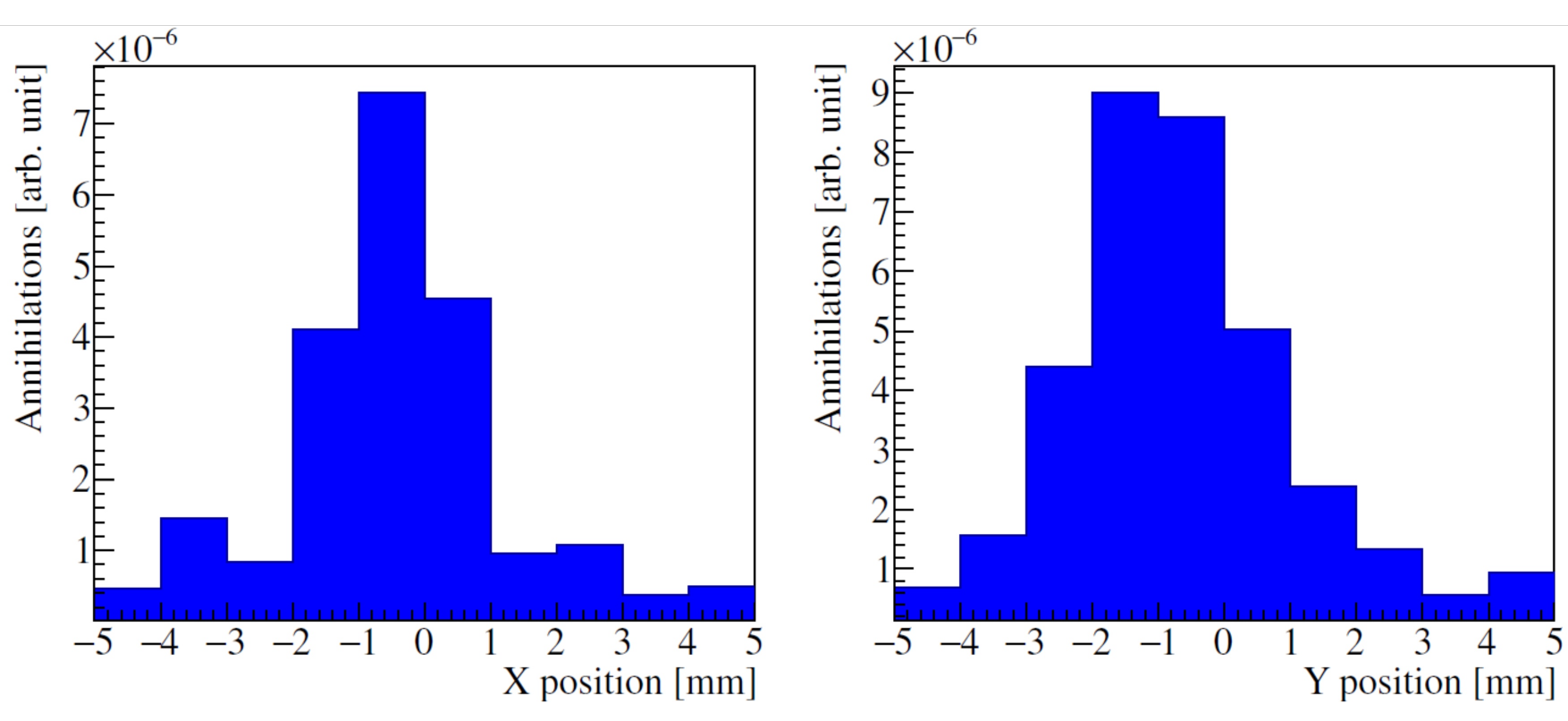}
\caption{
Beam profiles observed by the foils at the entrance of MUSASHI when 19~kV is applied to the drift tube.
The rms widths are $\sigma_x=1.0$~mm and $\sigma_y=1.6$~mm by Gaussian fitting.
}
\label{bpm2}
\end{figure}

In the following, we describe the procedure to accumulate antiprotons from ELENA 
in the MUSASHI trap using the drift tube.
Prior to antiproton injection, 1--3$\times10^8$ electrons are injected from the electron gun into the MUSASHI trap~\cite{Kuroda2005}. 
The electron plasma is then radially expanded ($>5$~mm) to increase the overlap with the incoming antiproton bunch.
Figure~\ref{TD_tstr} shows a typical time evolution of the cumulative annihilation counts 
measured by the two scintillator bars during one accumulation cycle.
The rise at $t=36$~s corresponds to antiproton injection.
The typical beam profiles observed by the Bo-PET foil detector at injection are shown in Fig.~\ref{bpm2}.
The root-mean-square widths are $\sigma_x=1.0$~mm horizontally and $\sigma_y=1.6$~mm vertically by Gaussian fitting 
which is small enough compared with a physical aperture of the MUSASHI trap electrodes of 20~mm.
The offset comes from alignment error of the foils against the trap axis which has been observed for years.
A negative high voltage of -12~kV is applied on the ring electrodes UCE and DCE (Fig.~\ref{exsetup})
to confine antiprotons decelerated by the foils.
A fraction of the high-energy antiprotons have annihilated at this time,
but only a few of them are counted due to pile-up.
Antiprotons are then electron-cooled for 40~s to an energy of less than 1~eV~\cite{Kuroda2014-2}.
The negative high voltage is then switched off at $t=76$~s.
Hot antiprotons (due to the poor overlap with electrons at high radius) annihilate at this time. 
The cooled antiprotons are extracted to the Bo-PET foils upstream of the MRE after $t=95$~s 
by electrostatic potential manipulations.
By slowly extracting the antiprotons for 10~s the annihilation rate does not saturate the detector.
Hence all annihilations are counted to estimate the number of trapped antiprotons.
The antiprotons are extracted downstream for antihydrogen production 
with a beam energy of 1--1000~eV.
Both a pulsed extraction of the antiprotons from the trap over a duration of $10^{-5}$~s 
or a slow extraction over several seconds are possible.

\begin{figure}[htb]
\centering
\includegraphics[clip,width=9cm]{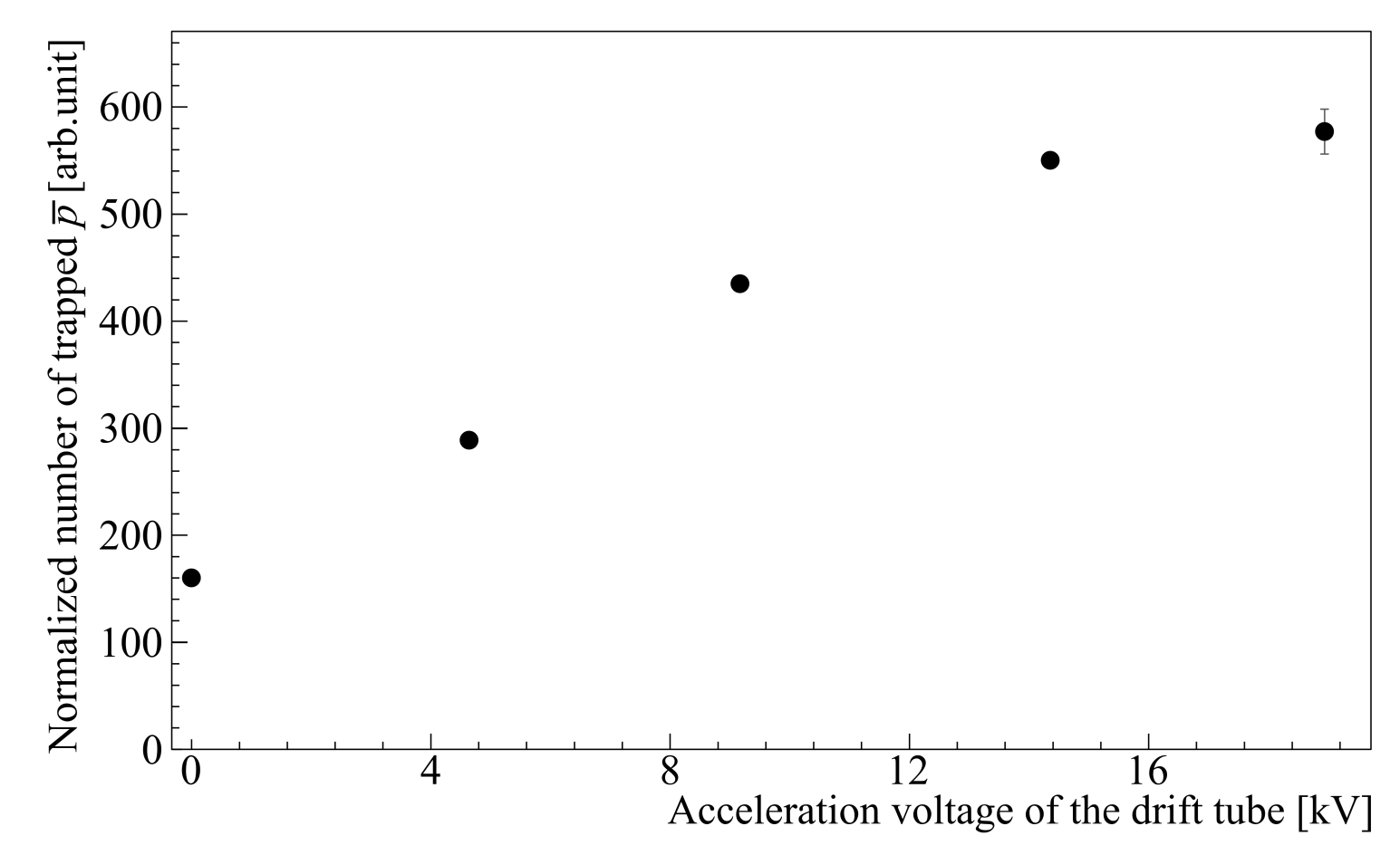}
\caption{The number of trapped antiprotons as a function of acceleration voltage of the drift tube,
normalized to the number of injected antiprotons from ELENA.
The error bars below 16~kV are smaller than the dots.}
\label{Trap_vs_DTHV}
\end{figure}

\begin{figure}[htb]
\centering
\includegraphics[clip,width=85mm]{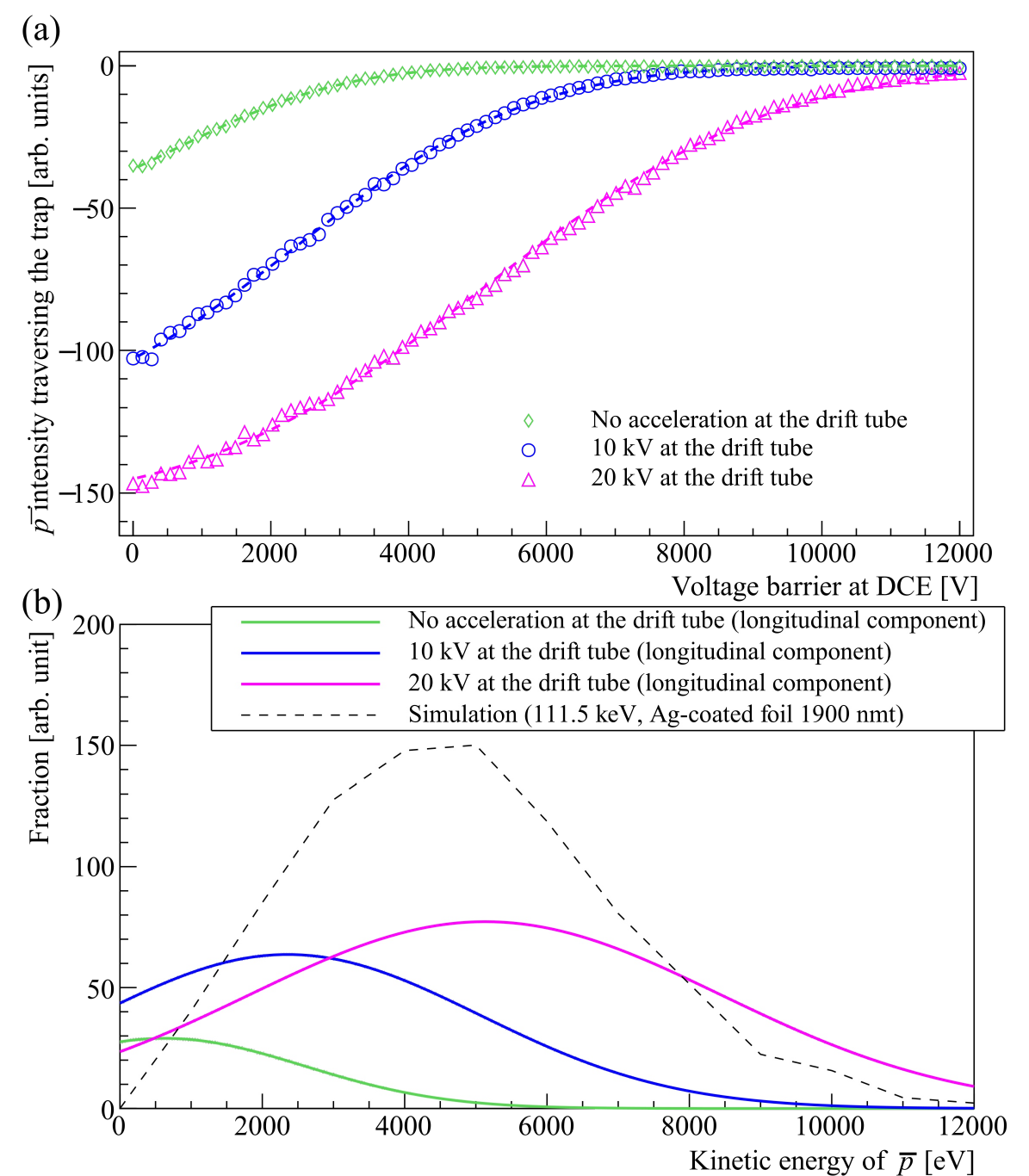}
\caption{
(a) The signal intensity of the antiproton beam that traversed the trap and annihilated in the beam dump
as a function of the voltage barrier applied to DCE for beam pulses 
accelerated by the drift tube by 0~kV, 10~kV, and 20~kV. 
(b) The longitudinal component of the energy distributions of transmitted antiprotons through the degrader foils.
The black dotted line shows the normalized simulation results~\cite{Nordlund2022}.
}
\label{edist}
\end{figure}

Figure~\ref{Trap_vs_DTHV} shows the number of trapped antiprotons observed by the two scintillator bars 
as a function of the acceleration voltage of the drift tube.
By applying 19~kV, the number is increased by a factor of four, compared to the case without acceleration. 
The detection efficiency of the two scintillator bars was estimated using the GEANT4 toolkit. 
In the simulation, events with higher energy deposit on scintillator bars than a threshold 
are counted when antiprotons are annihilated at the center of the MUSASHI trap, 
which is surrounded by a massive magnetic shield, SUS bore, superconducting coil, and cylindrical electrodes. 
The estimated detection efficiency is $0.50\pm0.05$\%, including possible error, 
primarily due to the foil position and annihilation position distribution uncertainties.
According to this detection efficiency,
the trapped number of antiprotons in MUSASHI corresponds to $(1.4\pm0.2)\times10^6$.
The number of antiprotons supplied from ELENA at the time of these experiments is about $(5.5\pm0.9)\times10^6$ per bunch every 2 minutes.
Thus, the efficiency is $(26\pm6)$\%.
With the previous setup of the AD and the RFQD, 
$3.5\times10^7$ antiprotons were supplied from the AD and
20\% of them were decelerated by the RFQD~\cite{Lombardi2001}.
They were injected into the MUSASHI trap and the typical trapped number was $1.5\times10^6$~\cite{Kuroda2012}.
The efficiency of the trapped number divided by the decelerated fraction was $1.5/(35\times0.2)=21$\%,
which is comparable to the value obtained with the drift tube.

\section{Discussion}
Figure~\ref{edist}(a) shows the signal intensity of the antiproton beam that traversed the trap and annihilated in the beam dump
as a function of the voltage barrier applied to DCE electrode of the MUSASHI, for beam pulses that were accelerated by the drift tube by 0~kV, 10~kV, and 20~kV. 
It is measured by integrating the signal of Cherenkov detector \#2 from $t=0$ to $t=10$~$\mu$s.
In the absence of a detailed physics analysis, we model the distribution of pbar energy due to scattering in the foil
as a Poisson process, which in the limit of large numbers gives a Gaussian.
We therefore model the signals in Fig.~\ref{edist}(a) using the integral of a gaussian, namely the error function.
Although the momentum spread $\Delta p/p\approx10^{-3}$~\cite{Bartmann2018} of the beam provided by ELENA 
is 10~times smaller compared to the beam of RFQD, 
the longitudinal energy of the transmitted antiprotons remained distributed over more than 10~keV~\cite{Kuroda2012}.

Molecular dynamics simulations in the recoil interaction approximation (MD-RIA) were recently carried out 
to study the transmission of antiprotons through degrader foils~\cite{Nordlund2022}. 
The simulations showed that nuclear scattering of antiprotons on atomic targets into large scattering angles 
occurs with a high probability, especially when the antiprotons are slowed down to keV-scale energies. 
This increases the effective path length of the antiprotons in the foils 
and causes a significant fraction to annihilate. 
In ref.\cite{Nordlund2022}, the simulation shows an example when antiprotons with a kinetic energy of 111.5~keV 
with an energy spread of 5~keV (standard deviation) traversed a 1900~nm thick Bo-PET foil 
with 25~nm thick Ag coatings on both surfaces.
Only $\approx40$\% of the antiprotons were predicted to emerge from the foil that may be trapped,
with a kinetic energy and angle corresponding to a Larmor radius of less than 5~mm.
Further rejecting the transmitted antiprotons, that emerged with an angle larger than 50~degrees 
relative to the normal of the foil surface, further reduced the trappable fraction to $\approx$30\%. 
These values roughly agree with the experimental trapping efficiency of 26\% 
within the uncertainties of the thicknesses of the foils, 
despite the fact that the simulation does not include the losses that practically occur in the trap after capture. 
The beam energy of 111.5~keV used in these simulations corresponds to the estimated mean value 
utilized in the previous experiment reported in ref.\cite{Kuroda2012}.
The energy distribution of the transmitted antiprotons (indicated by the dotted line in Fig.~\ref{edist}) 
shows almost no particles at $<1$~keV, as nuclear scattering and atomic capture cause them to annihilate. 
In these simulations, the processes that occur at such low energies are not fully understood.
The simulated and experimental distributions cannot be directly compared since they correspond to the total 
and longitudinal kinetic energies, respectively. 
The experiment may also be affected by pinholes that developed on the surfaces of the foils.

There are several possible reasons for the broad energy distribution which was observed.
First, the temporal ringing of the high voltage pulse applied to the drift tube (Fig.~\ref{oscillo_pulse}). 
It caused modulations in the energy of the antiproton pulse prior to its arrival at the degrader foils. 
The timing of the high voltage pulse was optimized by varying it in 50~ns steps so that the number of trapped antiprotons was maximized.
The timing jitter of this trigger signal provided by the ELENA control system was about 10~ns relative to the arrival of the antiprotons~\cite{Chohan2014}. 
Assuming that the antiprotons are distributed within the pulse following a Gaussian distribution, 
we estimate that some 15\% of the antiprotons may not have reached the nominal acceleration energy. 
Second, MD-RIA simulations~\cite{Nordlund2022} show that when a mono-energetic antiproton beam at 111.5~keV traverses a Bo-PET foil with a uniform thickness of 1800~nm, 
nuclear scattering effects causes the transmitted antiprotons to acquire an energy spread of 5~keV (FWHM). 
Third, the simulations indicate that a $\pm100$~nm thickness variation over the surface of the Bo-PET foils leads to 8~keV energy spread, 
whereas our foils have a spatial thickness uniformity of not better than 10\%.  

The simulation also suggests that nuclear scattering in the antiproton energy range of a few keV 
is more significant when materials with a higher atomic number (such as silver) are used. 
This leads to a reduction in the number of transmitted antiprotons and 
to an expansion of the beam size after traversing the foils.
This reduces the fraction of antiprotons that overlap with pre-confined electrons in the trap,
so that the trapping efficiency deteriorates.
The simulation suggests that the transmitted fraction would be higher 
if a light material such as aluminum were used instead of the silver electrodes~\cite{Nordlund2022}.
In the future, we will replace the silver strips with aluminum ones, thereby improving the trapping efficiency.

\section{Summary}
The ASACUSA collaboration has developed a drift tube accelerator to optimize the injection energy 
of antiprotons into the degrader foils at the entrance of the MUSASHI trap,
which is an efficient solution to work with ELENA supplying antiprotons at a fixed energy of 100~keV.
Commissioning has been performed, which shows that the drift tube operated successfully 
and that $(1.4\pm0.2)\times10^6$ antiprotons per bunch are confined by biasing the drift tube at 19~kV
which corresponds to a trapping efficiency of $(26\pm6)$\%.
The trapping efficiency may be improved by replacing the silver strips on the surface of the degrader foils 
with aluminum, as suggested by MD-RIA simulations.

\section*{Acknowledgments}
The authors are grateful to the AD and ELENA groups for their contributions on the smooth 
and successful operations of ELENA after the long shutdown 2 at CERN.
This work is supported by JSPS KAKENHI Fostering Joint International Research No.~B~19KK0075, 
Grant-in-Aid for Scientific Research No.~B~20H01930; 
the Austrian Science Fund (FWF) Grant Nos. P 32468, W1252-N27, and P 34438; 
Special Research Projects for Basic Science of RIKEN; 
Universit\`{a} di Brescia and Istituto Nazionale di Fisica Nucleare; 
the European Union's Horizon 2020 research and innovation program 
under the Marie Sklodowska-Curie Grant Agreement No. 721559;
Heisenberg Program and Individual Grant Program of the Deutsche Forschungsgemeinschaft;
and Research Grants Program of the Royal Society.


\begin{thebibliography}{26}
\bibitem{Maury1997}
	S.~Maury, ``The Antiproton Decelerator: AD,'' 
	\href{https://doi.org/10.1023/A:1012632812327}{\emph{Hyperfine Interact.} 109, 43-52 (1997).}
\bibitem{Bartmann2018}
	W.~Bartmann {\it et al.}, ``The ELENA facility,''
	\href{https://doi.org/10.1098/rsta.2017.0266}{\emph{Phil. Trans. R. Soc. A} 376, 20170266 (2018).}
\bibitem{Ahmadi2017}
	M.~Ahmadi {\it et al.}, ``Observation of the hyperfine spectrum of antihydrogen,''
	\href{https://doi.org/10.1038/nature23446}{\emph{Nature} 548, 66-69 (2017).}
\bibitem{Ahmadi2018}
	M.~Ahmadi {\it et al.}, ``Characterization of the 1S-2S transition in antihydrogen,''
	\href{https://doi.org/10.1038/s41586-018-0017-2}{\emph{Nature} 557, 71-75 (2018).}
\bibitem{ALPHA2020}
	The ALPHA Collaboration, ``Investigation of the fine structure of antihydrogen,''
	\href{https://doi.org/10.1038/s41586-020-2006-5}{\emph{Nature} 578, 375-380 (2020).}
\bibitem{Eberhard2019}
	E.~Widmann {\it et al.}, ``Hyperfine spectroscopy of hydrogen and antihydrogen in ASACUSA,''
	\href{https://doi.org/10.1007/s10751-018-1536-9}{\emph{Hyperfine Interact.} 240, 5 (2019).}
\bibitem{Gabrielse2012}
	G.~Gabrielse {\it et al.}, ``Trapped Antihydrogen in Its Ground State,''
	\href{https://doi.org/10.1103/PhysRevLett.108.113002}{\emph{Phys. Rev. Lett.} 108, 113002 (2012).}
\bibitem{Hori2016}
	M.~Hori {\it et al.}, ``Buffer-gas cooling of antiprotonic helium to 1.5 to 1.7~K, and antiproton-to-electron mass ratio,''
	\href{https://doi.org/10.1126/science.aaf6702}{\emph{Science} 354, 610-614 (2016).}
\bibitem{Borchert2022}
	M.J.~Borchert {\it et al.}, ``A 16-parts-per-trillion measurement of the antiproton-to-proton charge-mass ratio,''
	\href{https://doi.org/10.1038/s41586-021-04203-w}{\emph{Nature} 601, 53-57 (2022).}
\bibitem{DiSciacca2013}
	J.~DiSciacca {\it et al.}, ``One-Particle Measurement of the Antiproton Magnetic Moment,''
	\href{http://dx.doi.org/10.1103/PhysRevLett.110.130801}{\emph{Phys. Rev. Lett.} 110, 130801 (2013).}
\bibitem{Kellerbauer2008}
	A.~Kellerbauer {\it et al.}, ``Proposed antimatter gravity measurement with an antihydrogen beam,''
	\href{https://doi.org/10.1016/j.nimb.2007.12.010}{\emph{Nucl. Instrum. Methods Phys. Res. B} 266, 351-356 (2008).}
\bibitem{Perez2015}
	P.~Perez {\it et al.}, ``The gbar antimatter gravity experiment,''
	\href{https://doi.org/10.1007/s10751-015-1154-8}{\emph{Hyperfine Interact.} 233, 21-27 (2015).}
\bibitem{Bertsche2018}
	W.~A. Bertsche {\it et al.}, ``Prospects for comparison of matter and antimatter gravitation with alpha-g,''
	\href{https://doi.org/10.1098/rsta.2017.0265}{\emph{Philos. Trans. R. Soc. A} 376, 20170265 (2018).}
\bibitem{Anderson2023}
	E.~K.~Anderson {\it et al.}, ``Observation of the effect of gravity on the motion of antimatter,''
	\href{https://doi.org/10.1038/s41586-023-06527-1}{\emph{Nature} 621, 716-722 (2023).}
\bibitem{Kirchner2011}
	T.~Kirchner and H.~Knudsen, ``Current status of antiproton impact ionization of atoms and molecules: theoretical and experimental perspectives,''
	\href{http://dx.doi.org/10.1088/0953-4075/44/12/122001}{\emph{J. Phys. B: At. Mol. Opt. Phys.} 44, 122001 (2011).} 
\bibitem{Amsler2019}
	C.~Amsler {\it et al.}, ``ASACUSA proposal for ELENA,''
	\href{https://cds.cern.ch/record/2691506}{CERN-SPSC-2019-035 / SPSC-P-307-ADD-2 (2019).} 
\bibitem{Aumann2022}
	T.~Aumann {\it et al.}, ``PUMA, antiProton unstable matter annihilation,''
	\href{https://doi.org/10.1140/epja/s10050-022-00713-x}{\emph{Eur. Phys. J. A} 58, 88 (2022).}
\bibitem{Aghai2018}
	H.~Aghai-Khozani {\it et al.}, ``Measurement of the antiproton-nucleus annihilation cross-section at low energy,''
	\href{https://doi.org/10.1016/j.nuclphysa.2018.01.001}{\emph{Nucl. Phys. A} 970, 366-378 (2018).}
\bibitem{Aghai2021}
	H.~Aghai-Khozani {\it et al.}, ``Limits on antiproton-nuclei annihilation cross sections at $\sim125$ keV,''
	\href{https://doi.org/10.1016/j.nuclphysa.2021.122170}{\emph{Nucl. Phys. A} 1009, 122170  (2021).}
\bibitem{Kuroda2012}
	N.~Kuroda {\it et al.}, ``Development of a monoenergetic ultraslow antiproton beam source for high-precision investigation,''
	\href{https://doi.org/10.1103/PhysRevSTAB.15.024702}{\emph{Phys. Rev. ST Accel. Beams} 15, 024702 (2012).}
\bibitem{Lombardi2001}
	A.~M.~Lombardi, W.~Pirkl and Y.~Bylinsky, ``First operating experience with the CERN decelerating RFQ for antiprotons,'' 
	\href{https://doi.org/10.1109/PAC.2001.987575}{PACS2001. Proceedings of the 2001 Particle Accelerator Conference (Cat. No.01CH37268), Chicago, IL, USA, pp. 585-587 vol.1 (2001).}
\bibitem{Mohri2003}
	A.~Mohri, Y.~Yamazaki, ``A possible new scheme to synthesize antihydrogen and to prepare a polarised antihydrogen beam,''
	\href{https://doi.org/10.1209/epl/i2003-00509-0}{\emph{Europhys. Lett.} 63, 207 (2003).}
\bibitem{Enomoto2010}
	Y.~Enomoto {\it et al.}, ``Synthesis of cold antihydrogen in a cusp trap,''
	\href{https://doi.org/10.1103/PhysRevLett.105.243401}{\emph{Phys. Rev. Lett.} 105, 243401 (2010).}
\bibitem{Kuroda2014}
	N.~Kuroda {\it et al.}, ``A source of antihydrogen for in-flight hyperfine spectroscopy,''
	\href{https://doi.org/10.1038/ncomms4089}{\emph{Nat. Commun.} 5, 3089 (2014).}
\bibitem{Kolbinger2021}
	B.~Kolbinger {\it et al.}, ``Measurement of the principal quantum number distribution in a beam of antihydrogen atoms,''
	\href{https://doi.org/10.1140/epjd/s10053-021-00101-y}{\emph{Eur. Phys. J. D} 75, 91 (2021).}
\bibitem{Amsler2022}
	C.~Amsler {\it et al.}, ``Reducing the background temperature for cyclotron cooling in a cryogenic Penning-Malmberg trap,''
	\href{https://doi.org/10.1063/5.0093360}{\emph{Phys. Plasmas} 29, 083303 (2022).}
\bibitem{Hori2004}
	M.~Hori, ``Parallel plate chambers for monitoring the profiles of high-intensity pulsed antiproton beams,''
	\href{https://doi.org/10.1016/j.nima.2003.11.200}{\emph{Nucl. Instrum. Methods Phys. Res. A} 522, 420-431 (2004).}
\bibitem{Todoroki2016}
	K.~Todoroki {\it et al.}, ''Instrumentation for measurement of in-flight annihilations of 130 keV antiprotons on thin target foils,''
	\href{https://doi.org/10.1016/j.nima.2016.08.026}{\emph{Nucl. Instrum. Methods Phys. Res. A} 835, 110-118 (2016).}
\bibitem{Traykov2008}
	E.~Traykov {\it et al.}, ''Production and trapping of radioactive atoms at the TRI$\mu$P facility,''
	\href{https://doi.org/10.1016/j.nimb.2008.05.077}{\emph{Nucl. Instr. and Meth. in Phys. Res. B} 266, 4532-4536 (2008).}
\bibitem{Schury2017}
	P.~Schury {\it et al.}, ''Observation of doubly-charged ions of francium isotopes extracted from a gas cell,''
	\href{https://doi.org/10.1016/j.nimb.2017.06.014}{\emph{Nucl. Instr. and Meth. in Phys. Res. B} 407, 160-165 (2017).}
\bibitem{Husson2016}
	A.~Husson {\it et al.}, ''A pulsed high-voltage decelerator system to deliver low-energy antiprotons,''
	\href{https://doi.org/10.1016/j.nima.2021.165245}{\emph{Nucl. Instrum. Methods Phys. Res. A} 1002, 165245 (2021).}
\bibitem{Fischer2024}
	J.~Fischer {\it et al.}, ''Design and characterization of an antiproton deceleration beamline for the PUMA experiment,''
	\href{https://doi.org/10.1016/j.nimb.2024.165318}{\emph{Nucl. Instrum. Methods Phys. Res. B} 550, 165318 (2024).}
\bibitem{Kuroda2005}
	N.~Kuroda {\it et al.}, ``Confinement of a Large Number of Antiprotons and Production of an Ultraslow Antiproton Beam,''
	\href{https://doi.org/10.1103/PhysRevLett.94.023401}{\emph{Phys. Rev. Lett.} 94, 023401 (2005).}
\bibitem{Mohri1998}
	A.~Mohri {\it et al.}, ``Confinement of Nonneutral Spheroidal Plasmas in Multi-Ring Electrode Traps,''
	\href{https://doi.org/10.1143/JJAP.37.664}{\emph{Jpn. J. Appl. Phys.} 37, 664 (1998).}
\bibitem{Hori2003}
	M.~Hori {\it et al.}, ``Analog Cherenkov detectors used in laser spectroscopy experiments on antiprotonic helium,''
	\href{http://dx.doi.org/10.1016/S0168-9002(02)01618-2}{\emph{Nucl. Instrum. Methods Phys. Res. A} 496, 102-122 (2003).}
\bibitem{Agostinelli2003}
	S. Agostinelli {\it et al.}, ``Geant4 -- a simulation toolkit,'' 
	\href{https://doi.org/10.1016/S0168-9002(03)01368-8}{\emph{Nucl. Instrum. Methods Phys. Res. A} 506, 250-303 (2003).}
\bibitem{Kuroda2014-2}
	N.~Kuroda {\it et al.}, ``First Observation of a (1, 0) Mode Frequency Shift of an Electron Plasma at Antiproton Beam Injection,''
	\href{https://doi.org/10.1103/PhysRevLett.113.025001}{\emph{Phys. Rev. Lett.} 113, 025001 (2014).}
\bibitem{Nordlund2022}
	K.~Nordlund {\it et al.}, ``Large nuclear scattering effects in antiproton transmission through polymer and metal-coated foils,''
	\href{http://dx.doi.org/10.1103/PhysRevA.106.012803}{\emph{Phys. Rev. A} 106, 012803 (2022).}
\bibitem{Chohan2014}
	V.~Chohan {\it et al.}, ``Extra Low ENergy Antiproton (ELENA) ring and its Transfer Lines: Design Report,''
	\href{https://cds.cern.ch/record/1694484}{CERN Yellow Reports: Monographs (2014).}
\end{thebibliography}
\end{document}